\documentclass{article}
\usepackage{dcase2020,amsmath,graphicx,url,times,booktabs, tabularx, verbatim}
\usepackage{cite, amsfonts}
\usepackage{subcaption}

\title{CAPTURING SCATTERED DISCRIMINATIVE INFORMATION USING\\A DEEP ARCHITECTURE IN ACOUSTIC SCENE CLASSIFICATION}

\name{Hye-jin Shim$^*$\thanks{$^*$Equal contribution},
       Jee-weon Jung$^*$
       Ju-ho Kim,
       Ha-jin Yu$^\dagger$\thanks{$^\dagger$Corresponding author}\thanks{This research was supported by Basic Science Research Program through the National Research Foundation of Korea(NRF) funded by the Ministry of Science, ICT \& Future Planning(2020R1A2C1007081).},
       }
\address{University of Seoul, School of Computer Science, Seoul, South Korea \\}

\begin{document}
\ninept
\maketitle
\begin{sloppy}
\begin{abstract}
Frequently misclassified pairs of classes that share many common acoustic properties exist in acoustic scene classification (ASC).
To distinguish such pairs of classes, trivial details scattered throughout the data could be vital clues. 
However, these details are less noticeable and are easily removed using conventional non-linear activations (e.g. ReLU). 
Furthermore, making design choices to emphasize trivial details can easily lead to overfitting if the system is not sufficiently generalized. 
In this study, based on the analysis of the ASC task's characteristics, we investigate various methods to capture discriminative information and simultaneously mitigate the overfitting problem. 
We adopt a max feature map method to replace conventional non-linear activations in a deep neural network, and therefore, we apply an element-wise comparison between different filters of a convolution layer's output. 
Two data augment methods and two deep architecture modules are further explored to reduce overfitting and sustain the system's discriminative power. 
Various experiments are conducted using the detection and classification of acoustic scenes and events 2020 task1-a dataset to validate the proposed methods. 
Our results show that the proposed system consistently outperforms the baseline, where the single best performing system has an accuracy of 70.4\% compared to 65.1\% of the baseline. 

\end{abstract}
\begin{keywords}
LCNN, CBAM, ASC, deep neural network
\end{keywords}

\section{Introduction}
\label{sec:intro}
The detection and classification of acoustic scene and events (DCASE) community has been hosting multiple challenges to utilize sound event information generated in everyday environment and physical events \cite{DCASE2018Workshop, DCASE2019Workshop, mcdonnell2020acoustic}.
DCASE challenges provide not only the dataset for various audio-related tasks, but also a platform to compare and analyze the proposed systems.
Among many kinds of tasks covered in DCASE challenges, acoustic scene classification (ASC) is a multi-class classification task that classifies an input recording into one of the predefined scenes. 

In the process of developing an ASC system, two major issues have been widely explored in recent research literature. 
One is the generalization of the system in domain mismatch conditions that could arise from different recording devices \cite{gharib2018unsupervised, primus2019acoustic, kosmider2019calibrating}. 
More specifically, if an ASC system is not generalized towards unknown devices, performance on different devices degrades in the test phase. 
Another critical issue is the occurrence of frequently misclassified classes (e.g. shopping mall - airport, tram - metro) \cite{heo2019acoustic, Jung2019}. 
Many acoustic characteristics coincide with these pairs of classes. 
Trivial details can be decisive clues for accurate classification; however, focusing on such details easily leads to a trade-off, thereby degrading generalization. 
In particular, due to the characteristics of the ASC task (see Section \ref{sec:char_ASC}), discriminative information is scattered rather throughout the recording. 
However, widely used convolutional neural network (CNN)-based models that exploit the ReLU activation function make feature representations sparse as it may discard negative values \cite{Wu2015ALC}.

\begin{figure}[!t]
  \begin{center}
    \centering
      \subcaptionbox{Devices\label{tsne_dev}}
        {\includegraphics[width=0.40\linewidth]{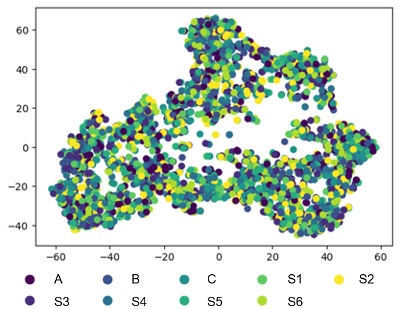}}
    \subcaptionbox{Scenes\label{tsne_scene}}
        {\includegraphics[width=0.40\linewidth]{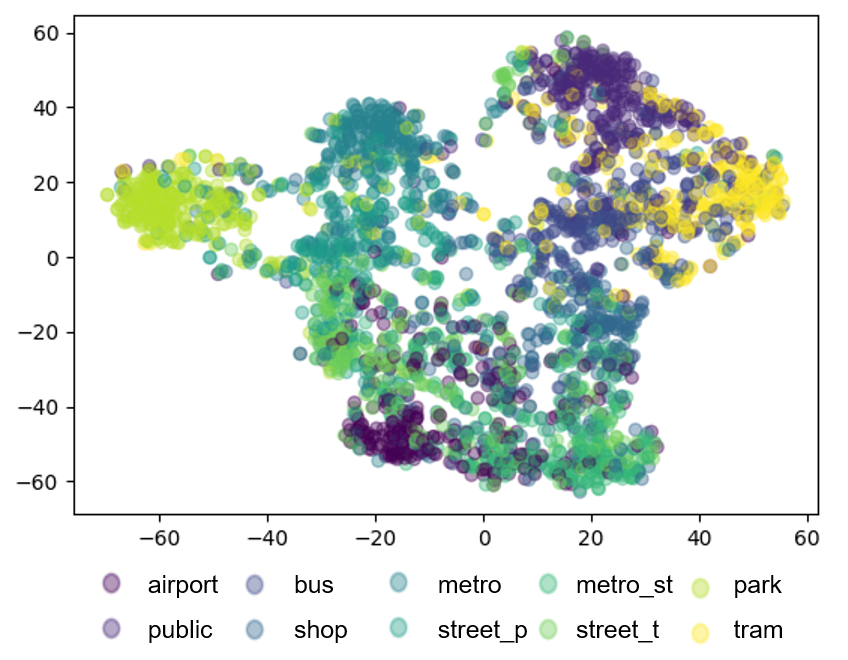}}
    \vskip -10pt
    \caption{t-SNE visualization results of embeddings. }
    \vskip -30pt
    \label{fig:tsne}
  \end{center}
\end{figure}

To investigate the aforementioned problems, we present a visualization of the representation vectors (i.e. embeddings, codes) of the baseline using a t-SNE algorithm \cite{maaten2008visualizing}, depicted in Figure \ref{fig:tsne}. 
Here, (a) and (b) refer to the result of plotted embeddings where different colors denote different device and scene labels, respectively. 
Figure \ref{fig:tsne}-(a) shows that the devices do not form noticeable clusters, indicating good generalization. 
However, it can be seen in Figure \ref{fig:tsne}-(b) that each scene does not have a clear decision boundary. 
Therefore, on leveraging this analysis, we focus on mitigating the misclassified classes. 

In this study, we explore several methods to reduce the removal of information and overfitting based on the characteristics of the ASC task, and this analysis is presented in Section \ref{sec:char_ASC}. 
Firstly, instead of common CNNs, we utilize a light CNN (LCNN) architecture \cite{wu2018light}.
LCNN is an architecture that adopts a max feature map (MFM) operation instead of non-linear activation functions such as ReLU or tanh. 
LCNN demonstrates the state-of-the-art performance in spoofing detection for automatic speaker verification (i.e. audio spoofing detection) \cite{lavrentyeva2019stc, lai2019assert}. 
Second, to mitigate overfitting, data augmentation and attention-based deep architectural modules are explored. 
Two data augmentation techniques, mix-up and SpecAugment are also investigated \cite{zhang2017mixup, park2019specaugment}. 
The convolutional block attention module (CBAM) and squeeze and excitation (SE) networks are studied for enhancing the discriminative power using minimum additional parameters \cite{hu2018squeeze, woo2018cbam}.

\section{Characteristics of ASC}
\label{sec:char_ASC}
In this section, we present an analysis of the characteristics of the ASC task. 
We assume that the discriminative information for the ASC task included in an audio recording is scattered. 
Sound cues could occur either consistently or occasionally. 
For example, consistently occurring sound cues, such as a low degree of reverberation and the sound of the wind imply outdoor location. 
Various sound events such as chirping of birds and barking of dogs are also important cues, but they are impactive and short, and they may only occur in some recordings that are labeled as ``parks’’. 
Therefore, important cues can have multiple characteristics; they are not focused on specific parts of the data, and they occur irregularly.
In our analysis, gathering scattered information that resides in an input recording is of interest. 

In tasks such as speaker and image classification, the target information in data is relatively clear. 
As speaker classification utilizes human voice to identify speaker identity, the discriminative information is concentrated in human voice rather than in non-speech segments. 
Therefore, many studies on speaker classification attempt to remove non-speech segments using techniques such as voice activity detection (VAD).
Similarly, many tasks in the image domains adopt various methods to focus only on the target object. 
Because of the differences in these tasks such as speaker and image classification versus the ASC task, we argue that different modeling approaches should be considered.

Audio spoofing detection is a task that shares similar characteristics with the ASC task considered in our analysis.
Audio spoofing detection also makes a binary decision whether an input utterance is spoofed. 
In the case of audio spoofing detection, discriminative information is more scattered because distortions occur in the entire audio file during the spoofing process.
Therefore, non-speech segments are also important because the distortion is not limited to the speech segments.
Previous studies also show that VAD could eliminate useful information \cite{lapidot2019effects, dinkel2017small}.
Considering these characteristics, and in order to not miss much information, it has been demonstrated that LCNN is particularly effective in audio spoofing detection \cite{lavrentyeva2019stc, lai2019assert}. 
This is because relatively less informative parts (i.e. negative values) could be removed using ReLU activation with a common CNN, making a sparse representation (as illustrated in Figure \ref{fig:mfm}-(a)). 
This phenomenon has been reported in \cite{wu2018light} to occur especially for the first few convolution layers.

We hypothesize that this phenomenon would apply to an ASC system too because the ASC task has commonalities in that important information is scattered across the data, similar to audio spoofing detection. 
To mitigate the problem of sparse representation in an ASC task, we propose to utilize MFM operation included in the LCNN architecture. 
As MFM operation selects feature maps with an element-wise competitive relationship, trivial information can be retained if the value is relatively high. 
Furthermore, focusing on trivial details could also lead to overfitting. 
Hence, in this study, we aim to adopt regularization methods, while introducing a minimum number of additional parameters and retaining the discriminative power of the system by applying state-of-the-art deep architecture modules.

\begin{figure}[!t]
  \begin{center}
    \centering
    \subcaptionbox{ReLU\label{cat}}
        {\raisebox{1cm}{\includegraphics[width=0.4\linewidth]{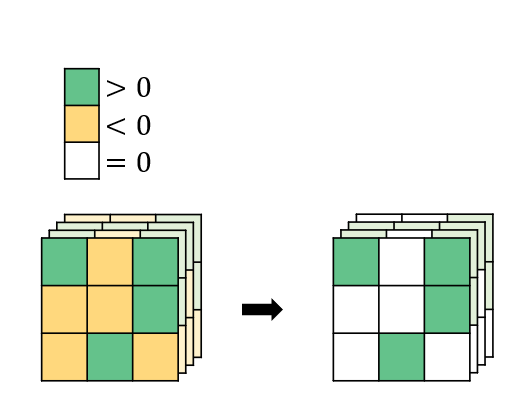}}}
      \subcaptionbox{MFM\label{relu}}
        {\includegraphics[width=0.5\linewidth]{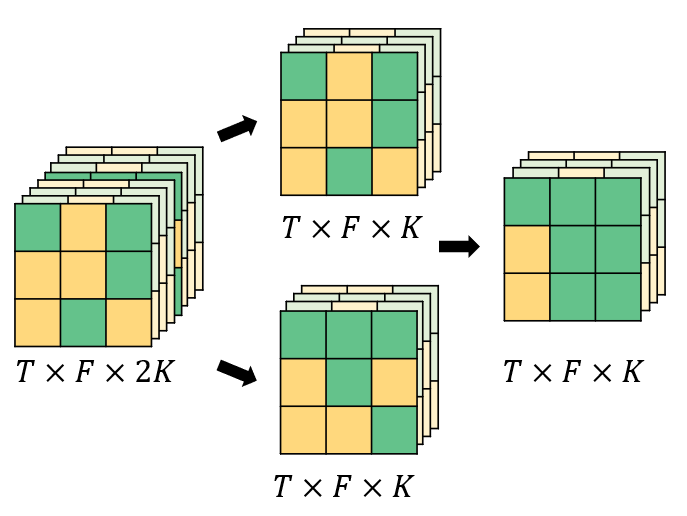}}
    \vskip -10pt
    \caption{Comparison of ReLU activation function (left) and MFM (right) . Orange, green, and white indicate negative, positive, and zero values, respectively. ReLU removes all negative values, while MFM considers the element-wise maximum one based on a comparative relationship}
    \label{fig:mfm}
    \vskip -10pt
  \end{center}
  \vskip -25pt
\end{figure}
\section{Proposed framework}
\label{sec:Proposed}

\begin{table}[t]
\caption{The LCNN architecture. The numbers in the output shape column refer to the frame (time), frequency, and the number of kernels. MFM, MaxPool and FC indicate max feature map, max pooling layer and fully-connected layer, respectively.} 
\centering 
\begin{tabular}{c c c} 
\toprule 
Type & Kernel/Stride & Output \\ [0.5ex] 
\hline 
Conv\_1 & {7 $\times$ 3 / 1 $\times$ 1 } & \textit{l} $\times$ 124 $\times$ 64  \\
MFM\_1 & - & \textit{l} $\times$ 124 $\times$ 32 \\
\hline
MaxPool\_1 & 2 $\times$ 2 / 2 $\times$ 2  & (\textit{l} / 2) $\times$ 62 $\times$ 32 \\
\hline
Conv\_2a & 1 $\times$ 1 / 1 $\times$ 1  & (\textit{l} / 2) $\times$ 62 $\times$ 64 \\
MFM\_2a & - & (\textit{l} / 2) $\times$ 62 $\times$ 32 \\
BatchNorm\_2a & -  & (\textit{l} / 2) $\times$ 62 $\times$ 32 \\
Conv\_2 & 3 $\times$ 3 / 1 $\times$ 1  & (\textit{l} / 2) $\times$ 62 $\times$ 96 \\
MFM\_2 & - & (\textit{l} / 2) $\times$ 62 $\times$ 48 \\
CBAM\_2 & - & (\textit{l} / 2) $\times$ 62 $\times$ 48 \\
\hline
MaxPool\_2 & 2 $\times$ 2 / 2 $\times$ 2  & (\textit{l} / 4) $\times$ 31 $\times$ 48 \\
BatchNorm\_2 & -  & (\textit{l} / 4) $\times$ 31 $\times$ 48 \\
\hline
Conv\_3a & 1 $\times$ 1 / 1 $\times$ 1  & (\textit{l} / 4) $\times$ 31 $\times$ 96 \\
MFM\_3a & - & (\textit{l} / 4) $\times$ 31 $\times$ 48 \\
BatchNorm\_3a & -  & (\textit{l} / 4) $\times$ 31 $\times$ 48 \\
Conv\_3 & 3 $\times$ 3 / 1 $\times$ 1  & (\textit{l} / 4) $\times$ 31 $\times$ 128 \\
MFM\_3 & - & (\textit{l} / 4) $\times$ 31 $\times$ 64 \\
CBAM\_3 & - & (\textit{l} / 4) $\times$ 31 $\times$ 64 \\
\hline
MaxPool\_3 & 2 $\times$ 2 / 2 $\times$ 2  & (\textit{l} / 8) $\times$ 16 $\times$ 64 \\
\hline
Conv\_4a & 1 $\times$ 1 / 1 $\times$ 1  & (\textit{l} / 8) $\times$ 16 $\times$ 128 \\
MFM\_4a & - & (\textit{l} / 8) $\times$ 16 $\times$ 64 \\
BatchNorm\_3a & -  & (\textit{l} / 8) $\times$ 16 $\times$ 64 \\

Conv\_4 & 3 $\times$ 3 / 1 $\times$ 1  & (\textit{l} / 8) $\times$ 16 $\times$ 64 \\
MFM\_4 & - & (\textit{l} / 8) $\times$ 16 $\times$ 32 \\
CBAM\_4 & - & (\textit{l} / 8) $\times$ 16 $\times$ 32 \\
\hline
BatchNorm\_4 & -  & (\textit{l} / 8) $\times$ 16 $\times$ 32 \\
\hline
Conv\_5a & 1 $\times$ 1 / 1 $\times$ 1  & (\textit{l} / 8) $\times$ 16 $\times$ 64 \\
MFM\_5a & - & (\textit{l} / 8) $\times$ 16 $\times$ 32 \\
BatchNorm\_5a & -  & (\textit{l} / 8) $\times$ 16 $\times$ 32 \\
Conv\_5 & 3 $\times$ 3 / 1 $\times$ 1  & (\textit{l} / 8) $\times$ 16 $\times$ 64 \\
MFM\_5 & - & (\textit{l} / 8) $\times$ 16 $\times$ 32 \\
CBAM\_5 & - & (\textit{l} / 8) $\times$ 16 $\times$ 32 \\
\hline
MaxPool\_5 & 2 $\times$ 2 / 2 $\times$ 2  & (\textit{l} / 16) $\times$ 8 $\times$ 32 \\
\hline
FC\_1 & - & 160 \\
MFM\_FC1 & - & 80 \\
\hline
FC\_2 & - & 10 \\
\bottomrule
\end{tabular}
\vskip -20pt
\label{tab:deeparc_lcnn} 
\end{table}

\subsection{LCNN}
\label{ssec:LCNN}
LCNN is a deep learning architecture, initially designed for face recognition with noisy labels \cite{wu2018light}. 
Its main feature is a novel operation referred to as max feature map (MFM) that replaces the non-linear activation function of a deep neural network (DNN). 
MFM operation extends the concept of maxout activation \cite{goodfellow2013maxout} and adopts a competitive scheme between filters of a given feature map. 
In this study, we introduce the MFM operation to the ASC task, based on two assumptions. To the best of our knowledge, this is the first report on such an implementation. 
Firstly, we hypothesize that scattered discriminative information can relatively reside throughout an input feature map, compared to widely used ReLU non-linearity that discards negative values.  
Secondly, we note that MFM operations demonstrate state-of-the-art performance in audio spoofing detection in which two tasks share common properties.

The implementation of an MFM operation can be denoted as follows. 
Let $a$ be a given feature map derived through a convolution layer, $a \in \mathbb{R}^{K \times T \times F}$, where $K$, $T$, and $F$ refer to the number of output channels, time domain frames, and frequency bins, respectively. 
We split $a$ into two feature maps, $a_1$ and $a_2$, $a_1$, $a_2$ $\in \mathbb{R}^{\frac{K}{2} \times T \times F}$. 
The MFM applied feature map is obtained by $Max(a_1, a_2)$, element-wise. 
Figure \ref{fig:mfm}-(b) illustrates the MFM operation.

Specifically, our design of LCNN is similar to that of \cite{lavrentyeva2019stc}, with some modifications.
The architecture of \cite{lavrentyeva2019stc} is also a modified version of the original LCNN \cite{wu2018light}, applying additional batch normalization used after a max pooling layer. 
Table \ref{tab:deeparc_lcnn} provides details of the architecture of the proposed system that adopts an LCNN.
Conv$\_$a, MFM$\_$a, BatchNorm, Conv, MFM, CBAM can be seen as a block, and 4 blocks are implemented to contain an adequate number of parameters.
The number of blocks is determined based on comparative experiments.

\subsection{Regularization and deep architecture modules}
\label{ssec:dataAug}
With limited labelled data and recent DNNs with many parameters, overfitting easily occurs in DNN-based ASC systems \cite{mcdonnell2020acoustic, Jung2019, mun2017generative, zhang2017mixup, park2019specaugment}. 
To account for overfitting, our design choices include data augmentation methods and deep architecture modules for generalization purposes with enhanced model capacity. 
For the regularization purpose, we adopt two data augmentation methods: mix-up \cite{zhang2017mixup} and specAugment \cite{park2019specaugment}. 
Let $x_i$ and $x_j$ be two audio recordings that belong to class $y_i$ and $y_j$, respectively, where $y$ is a one-hot vector. 
A mix-up operation creates an augmented audio recording with a corresponding soft-label using two different recordings. 
Formally, an augmented audio recording can be denoted as the following:
\begin{equation}
\begin{aligned}
    x' = \lambda x_i + (1 - \lambda) x_j,\\
    y' = \lambda y_i + (1- \lambda) y_j,
\end{aligned}
\end{equation}
where $\lambda$, is a random variable drawn from $Beta(\alpha, \alpha)$, and $\alpha \in (0, \inf)$, is a real value between 0 and 1. 
With a rather simple implementation, mix-up is widely adopted for the ASC task in the literature. 

We also adopt specAugment \cite{park2019specaugment}, which was first proposed for robust speech recognition that masks a certain region of two-dimensional input feature (i.e. spectrogram, Mel-filterbank energy). 
Among three methodologies proposed in the paper, we adopt frequency masking and time masking. 
Let $x$, $x \in$ $\mathbb{R}^{T \times F}$ be a Mel-filterbank energy feature extracted from an input audio recording, where $T$ and $F$ are the number of frames and Mel-frequency bins, respectively, and $t$ and $f$ are indices for $T$ and $F$ respectively. 
To apply time masking, we randomly select $t_{stt}$ and $t_{end}$, $t_{stt} \leq t_{end} \leq t_T$, where $stt$ and $end$ are indices for start and end, and then, mask with 0. 
To apply frequency masking, we randomly select $f_{stt}$ and $f_{end}$, $f_{stt} \leq f_{end} \leq f_F$, and then, mask with 0. 
In this study, we sequentially apply specAugment and mix-up for better generalization. 

To increase model capacity while introducing minimum number of additional parameters to the model, we investigate two recent deep architecture modules: SE \cite{hu2018squeeze} and CBAM \cite{woo2018cbam}. 
SE focuses on the relationship between different channels of a given feature map. 
SE first \textit{squeezes} the input feature map via a global average pooling layer to derive a channel descriptor that includes the global spatial (time and frequency in ASC) context. 
Then, using minimal number of additional parameters, SE re-calibrates channel-wise dependencies via an \textit{excitation} step. 
Specifically, the excitation step adopts two fully-connected layers that input a derived channel descriptor and output a re-calibrated channel descriptor. 
SE transforms the given feature map by multiplying the re-calibrated channel descriptor, where each value in the channel descriptor is broadcasted to conduct element-wise multiplication with each filter of a feature map. 
In our experiments that incorporate the SE module, we apply SE to the output of each residual block following the methods reported in the literature. 
Further details regarding the SE module can be found in \cite{hu2018squeeze}. 

CBAM is a deep architecture module that sequentially applies channel attention and spatial attention. 
To derive a channel attention map, CBAM applies global max and average pooling operations to the spatial domain.
It then uses two fully-connected layers. 
Channel attention is applied by element-wise multiplication of the input feature map with the channel attention map, where each value of the channel attention map is broadcasted to fit the spatial domain. 
To derive a spatial attention map, CBAM applies two global pooling operations to the channel domain and then adopts a convolution layer. 
Spatial attention is also applied by element-wise multiplication of the feature map after channel attention with a derived spatial attention map. 
In our experiments using the CBAM module, we apply it to the output of each residual block following the literature. 
Further details regarding the CBAM module can be found in \cite{woo2018cbam}.

\section{Experiments}
\label{sec:exp}
\subsection{Dataset}
\label{ssec:db}
We use the DCASE2020 task1-a dataset for all our experiments. 
It includes 23,040 audio recordings 44.1 kHz with a 24-bit resolution, where each recording has a duration of 10 s. 
The dataset contains audio recordings from three real devices (A, B, and C) and six augmented devices (S1-S6). 
Unless explicitly mentioned, all performances in this paper are reported using the official DCASE2020 fold 1 configuration, which assigns 13,965 recordings as the training set and 2,970 recordings as the test set. 

\subsection{Experimental configurations}
Mel-spectrograms with 128 Mel-filterbanks are used for all experiments where the number of FFT bins, window length, and shift size are set to 2,048, 40 ms, and 20 ms, respectively. 
During the training phase, we randomly select 250 consecutive frames (5 s) instead of using the whole recording. 
In the test phase, an audio recording is split into three overlapping sub-recordings (i.e. 0~5 s, 2.5~7.5 s, and 5~10 s), and the mean of the output layer is used to perform classification. 
This technique has been reported to mitigate overfitting in previous works \cite{heo2019acoustic, Jung2018}. 

We use an SGD optimizer with a batch size of 24.
The initial learning rate is set to 0.001 and scheduled with a warm restart of stochastic gradient descent \cite{loshchilov2016sgdr}. 
For a single system, we train the DNN in an end-to-end fashion. 
For the ensemble system, support vector machine (SVM) classifiers are employed. 
Further technical details are provided in our technical report for facilitating the reproduction of conducted experiments \cite{Shim2020}.

\section{Result analysis}
\label{sec:res}
\begin{table}[t]
  \caption{Baseline comparison with other systems. Classification accuracies reported  using DCASE2020 fold1 configuration.}
  \label{tab:base}
  \centering
  \begin{tabular}{lc}
    \toprule
    System & Acc (\%) \\
    \midrule
    DCASE2019 baseline \cite{DCASE2019Workshop} & 46.5\\
    DCASE2020 baseline \cite{Heittola2020} & 54.1\\
    \midrule
    \textbf{\textit{Ours-baseline}} & 65.3\\
    \bottomrule
  \end{tabular}
\end{table}

Table \ref{tab:base} compares the baseline of this study with the two official baselines of the DCASE community. 
The DCASE2019 baseline inputs log Mel-spectrograms and uses convolution and fully-connected layers. 
Further, the DCASE2020 baseline inputs L3 embeddings \cite{Cramer_2019_ICASSP} extracted from another DNN and uses fully-connected layers for classification. 
Our baseline uses Mel-spectrograms as inputs, and it uses convolution, batch normalization \cite{ioffe2015batch}, and Leaky ReLU \cite{maas2013rectifier} layers with residual connection \cite{he2016deep}, where a SE module exists after each residual block\footnote{Model architecture and accuracies per each device and scene is presented in our technical report for the DCASE2020 challenge.}. 
The results show that our baseline outperforms the DCASE2020 baseline by over 10\% in classification accuracy.

\begin{table}[t]
  \caption{Effect analysis of LCNN, data augmentation, and deep architecture modules. }
  \label{tab:proposed}
  \centering
  \begin{tabular}{lcc}
    \toprule
    System & Config & Acc (\%) \\
    \midrule
    ResNet & - & 65.1\\
    ResNet(baseline) & mix-up & 65.3\\
    ResNet & SpecAug & 66.7\\
    ResNet & mix-up+SpecAug & 67.3\\
    \midrule
    LCNN & - & 67.1\\
    LCNN & mix-up & 68.4\\
    LCNN & SpecAug & 69.2\\
    LCNN & mix-up+SpecAug & 69.4\\
    \midrule
    LCNN & SE & 68.0\\
    LCNN & CBAM & 68.3\\
    LCNN & SE+CBAM & 68.2\\
    \midrule
    LCNN & mix-up+SpecAug+SE & 69.8\\
    LCNN (submitted) & mix-up+SpecAug+CBAM & \textbf{70.4}\\
    \bottomrule
  \end{tabular}
\end{table}

Table \ref{tab:proposed} describes the effectiveness of the proposed approaches using LCNN, SE, and CBAM. 
It also provides a comparison of the effects of using mix-up or/and specAugment data augmentation methods. 
Firstly, for comparing the system architecture without any data augmentation and deep architecture modules, ResNet and LCNN achieve accuracies of 65.1\% and 67.1\%, respectively.
To optimize the LCNN system, we also adjust the number of blocks and find that the original LCNN with 4 blocks achieves the best performance. 
Secondly, we validate the effectiveness of data augmentation. 
The results show that mix-up and specAugment are both effective and using a combination of these two methods is the best choice. 
Thirdly, we apply the deep architecture modules of SE and CBAM. 
From the results of the experiment, we observe that the CBAM is slightly better than SE. 

Table \ref{tab:conf} represents the results of comparing the frequently misclassified pair of two classes through a confusion matrix. 
Due to limited space, we omit the entire confusion matrix, instead, we have depicted only the top-5 frequently misclassified pairs. 
Except for the pair of shopping mall and street pedestrian, misclassified errors are reduced. 
There are several improvements for other misclassified pairs, but even in the top-5, we found that the total misclassified pair improved by 17\% compared to the baseline.

Table \ref{tab:eval} shows performances of the proposed systems, submitted for the DCASE2020 challenge task-1a. 
Our method comprise a 4-fold cross validation and apply a score-sum ensemble. 
For the ensemble, an SVM classifier using a kernel with a radial basis function is used. 
Ours-LCNN shows the result of our submitted LCNN system in which the system outperforms the baseline by over 15\% using less than one-fifth the number of parameters. 
Further, using a score-sum ensemble with another ASC system using audio tagging\footnote{Also submitted to DCASE2020 workshop; authors will add a citation if accepted.}, classification accuracy increased to 71.7\%.

\begin{table}[t]
  \caption{Comparison results of the number of frequently misclassified pairs of acoustic scenes between baseline and proposed system. Reduction refers to the number of confusion pairs between the two classes. 
  }
  \label{tab:conf}
  \centering
  \begin{tabular}{lccc}
    \toprule
    Class & Baseline & Proposed & Reduction\\
    \midrule
    Metro - Tram & 114 & 81 & \textbf{33}\\
    Shopping - Airport & 107 & 101 & \textbf{6}\\
    Shopping - Metro$\_$st & 84 & 56 & \textbf{28}\\ 
    Shopping - Street$\_$ped & 83 & 88 & -5\\ 
    Public$\_$square - Street$\_$ped & 74 & 70 & \textbf{4}\\ 
    \midrule
    Total & 462 & 396 & 66\\
    \bottomrule
  \end{tabular}
\end{table}

\begin{table}[t]
  \caption{Results of our submitted systems for the DCASE2020 challenge task1-a.}
  \label{tab:eval}
  \centering
  \begin{tabular}{lcc}
    \toprule
    System & \# Param & Acc (\%) \\
    \midrule
    DCASE2020 baseline \cite{Heittola2020} & 5M & 51.4\\
    \midrule
    Ours-LCNN & 0.9M & 68.5\\
    Ours-LCNN+tagging & 1.6M & 71.7\\
    \bottomrule
  \end{tabular}
\end{table}

\section{Conclusion}
\label{sec:conclusion}
In this paper, we assumed that the information that enables classification between different scenes with similar characteristics is scattered throughout the recording of the ASC task. 
In the case of a shopping mall and an airport, there was a common characteristic that they were reverberant and there was a babel of voices as they are indoors. 
Therefore, trivial details could be important cues to distinguish the two classes.
Based on this hypothesis, we proposed a method that is expected to capture this discriminative information better. 
We applied two deep architecture modules of LCNN and CBAM and two data augmentation methods of mix-up and specAugment. 
The proposed method helped to improve the system performance with less computation and overhead parameters.
We achieved an accuracy of 70.4\% using the single best performing system, compared to 65.1\% of the baseline.

\clearpage
\bibliographystyle{IEEEtran}
\bibliography{refs}
\end{sloppy}
\end{document}